\documentstyle[twoside,fleqn,espcrc2]{article}

\def\ln{\ell{n}}

\newcommand{\AmS}{{\protect\the\textfont2
  A\kern-.1667em\lower.5ex\hbox{M}\kern-.125emS}}

\title{Undoubled Chiral Fermions on a Lattice}

\author{
She-Sheng Xue
\address{
INFN - Milan Section, Via Celoria 16, Milan, Italy}}
       
\begin{document}

\begin{abstract}
We analyze the dynamics of an $SU_L(2)\otimes U_R(1)$ chiral theory on the 
lattice with a strong multifermion coupling. It is shown that no
spontaneous symmetry breaking occurs; the ``spectator'' fermion $\psi_R$ is
a free mode; doublers are decoupled as massive Dirac fermions consistently with
the chiral symmetries. In 1+1 dimension, we show that the right-handed 
three-fermion state disappears at the threshold and an undoubled left-handed
chiral fermion remains in the continuum limit.
 
\end{abstract}

\maketitle 

Since the demonstration of ``no-go'' theorem \cite{nn81} of Nielsen and 
Ninomiya in 1981 the problem of chiral fermion ``doubling'' and
``vector-like'' phenomenon on a lattice still exists if one insists on
preserving chiral symmetry. One of the ideas to get around this ``no-go''
theorem was proposed with a multifermion coupling model \cite{ep}. However, 
it was pointed out \cite{gpr} that such a model of
multifermion couplings fails to give chiral fermions in the continuum limit. 

We have been studying multifermion couplings on the lattice for years
\cite{xue91} and we believe that such models still have a chance to work \cite{xue96}.
Let us consider the following fermion action of the $SU_L(2)\otimes U_R(1)$
chiral symmetries on the lattice with a multifermion coupling. 
\begin{eqnarray}
S\!&\!=\!&\!S_k\!+\! g_2\bar\psi^i_L(x)\!\cdot\!\Delta\psi_R
\Delta\bar\psi_R\!\cdot\!\psi_L^i(x),\label{action}\\
S_k\!&\!=\!&\!{1\over 2a}\Big(\bar\psi^i_L 
\partial\psi^i_L\!+\!\bar\psi_R\partial\psi_R\Big)\nonumber
\end{eqnarray}
where ``$a$'' is the lattice spacing; $\psi^i_L$ ($i=1,2$) is an $SU_L(2)$
doublet, $\psi_R$ is an $SU_L(2)$ singlet and both are two-component
Weyl fermions. The $\psi_R$ is treated as a ``spectator'' fermion. The 
$\Delta(x)$ is a second order differential operator on the lattice.
The multifermion coupling $g_2$ is a dimension-10
operator relevant only for doublers $p=\tilde p+\pi_A$, but irrelevant for the 
chiral fermions $p=\tilde p\sim 0$ of the
$\psi^i_L$ and $\psi_R$. Action (\ref{action}) preserves the global chiral
symmetry $SU_L(2)\otimes U_R(1)$ and the $\psi^i_L(x)$ can be gauged to have
the exact local $SU_L(2)$ chiral gauge symmetry. In addition, action
(\ref{action}) possesses a $\psi_R$-shift-symmetry\cite{gp}, 
\begin{equation}
\psi_R(x) \rightarrow \psi_R(x)+\epsilon.
\label{shift}
\end{equation}
The $U_L(1)$ global symmetry relating to the conservation of 
the fermion number of the $\psi_L^i(x)$ is explicit in eq.(\ref{action}).

To seek a possible segment $g_2\gg 1$, where an undoubled
$SU_L(2)$-chiral fermion exists in the continuum limit, we are bound to
demonstrate the following properties of the theory in this segment: 
\begin{itemize}
\begin{enumerate}
\leftmargini=1.5em
\item the normal mode of $\psi_R$ is a free mode and decoupled; 

\item no spontaneous chiral symmetry breaking occurs ;

\item all doublers $p=\tilde p+\pi_A$ are bound to be massive
Dirac fermions and decoupled consistently with the $SU_L(2)\otimes U_R(1)$
chiral symmetry;

\item an undoubled chiral fermion $p=\tilde p$ of $\psi^i_L$ 
exists in the low-energy spectrum. 

\end{enumerate}
\end{itemize}
To prove the first and second points, we use the Ward identity stemming
from the $\psi_R$-shift-symmetry (\ref{shift}).
Considering the generating functional $W(\eta)$ of the theory, 
\begin{eqnarray}
W(\eta)&=&-\ln Z(\eta),
\label{part}\\
Z(\eta)&=&\int [d\psi_L^i d\psi_R]\exp\Big(-S+\bar\psi\eta\Big),\nonumber
\end{eqnarray}
we define the effective action
$\Gamma(\psi'^i_L,\psi_R')$ as
the Legendre transform of $W(\eta)$.
Making the parameter $\epsilon$ in eq.(\ref{shift}) to be space-time
dependent, and varying the generating functional (\ref{part}) according to the
transformation rule (\ref{shift}) for arbitrary $\epsilon(x)\not= 0$, we arrive
at the Ward identity
corresponding to the $\psi_R$-shift-symmetry of the action (\ref{action}): 
\begin{equation}
{1\over 2a}\partial\psi^\prime_R\!+\!g_2\!\langle\Delta\!\left(
\bar\psi^i_L\!\cdot\!
\Delta\psi_R\psi_L^i\right)\rangle\!-\!{\delta\Gamma\over\delta\bar
\psi'_R}=0.
\label{w}
\end{equation}
From this Ward identity, one can obtain all one-particle irreducible
vertices $\Gamma^{(n)}_R$ containing at least one external $\psi_R$. 
Taking functional derivatives of eq.(\ref{w}) with respect to appropriate
``primed'' fields and then putting external sources 
$\eta=0$, one can derive:
\begin{eqnarray}
\int_xe^{-ipx}
{\delta^{(2)}\Gamma\over\delta\psi'_R(x)\delta\bar\psi'_R(0)}&=&{i\over a}
\gamma_\mu\sin p^\mu,\label{free}\\
{1\over2}\Sigma^i(p)&\sim& w(p),
\label{ws2}
\end{eqnarray}
where $w(p)$ is the Fourier transform of 
${1\over2}\Delta(x)$, which is the Wilson factor \cite{wilson}. 
One has
\begin{equation}
w(p)=\sum_\mu(1-\cos p_\mu),\hskip0.2cm \Sigma^i(0)=0.
\label{zero}
\end{equation}
Eq.(\ref{free}) indicates an absence of the wave-function
renormalization of the $\psi_R$. Eqs.(\ref{ws2},\ref{zero}) show the
vanishing of the NJL symmetry breaking \cite{njl} for $p=0$. 

To adopt the technique of strong-coupling expansion in powers of
${1\over g_2}$ and we make a rescaling of the fermion fields 
\begin{equation}
\psi^i_L\rightarrow (g_2)^{1\over4}\psi^i_L;
\hskip0.2cm\psi_R\rightarrow (g_2)^{1\over4}\psi_R.
\label{rescale1}
\end{equation}
For the strong coupling $g_2\rightarrow\infty$, the kinetic
terms $S_k$ can be dropped and we compute the integral of 
$e^{-S(x)}$ in this limit: 
\begin{equation}
Z=2^{4N}\left(\det\Delta^2(x)\right)^4,
\label{stronglimit1}
\end{equation}
where the determinant is taken only over the lattice-space-time and ``$N$'' is
the number of lattice sites. 
For the non-zero eigenvalues of the operator
$\Delta^2(x)$, eq.~(\ref{stronglimit1}) shows an existence of the sensible 
strong-coupling limit. However, as for the zero eigenvalue of the operator
$\Delta(x)$, this strong-coupling limit is not analytic and the 
strong-coupling expansion in powers of ${1\over g_2}$ breaks down. 

To show the second point concerning the NJL symmetry breaking,
we calculate the two-point functions: 
\begin{eqnarray}
S^j_{RL}(x)&\equiv&\langle\psi_R(0),\bar\psi^j_L(x)\rangle,\label{srl}\\
S^j_{MR}(x)&\equiv&\langle\psi_R(0),[\bar\psi^j_L\cdot\psi_R]
\bar\psi_R(x)\rangle.
\label{smr}
\end{eqnarray}
At non-trivial leading order $O({1\over g_2})$, we get the recursion relations:
\begin{eqnarray}
S^j_{RL}(x)\!&\!=\!&\!{\sum^\dagger_\mu S^j_{MR}(x\!+\!\mu)\gamma_\mu
\over 8a^3g_2\Delta^2(x)},\label{re5}\\
S^j_{MR}(x)\!&\!=\!&\!{\sum^\dagger_\mu S^j_{RL}(x\!+\!\mu)\gamma_\mu
\over 2ag_2\Delta^2(x)}.
\label{re6}
\end{eqnarray}
These recursion relations are not valid where the operator $\Delta(x)$ has zero
eigenvalues. For $p\not=0$ and $\Delta(p)=2w(p)\not=0$, the Fourier transform of
these recursion relations leads to 
the solution ($p\not=0$), 
\begin{equation}
S^j_{RL}(p)\equiv \Sigma^j(p)=0,\hskip0.5cm
S^j_{MR}(p)=0.
\end{equation}
Together with eq.(\ref{zero}), we prove the second point.

We turn to the third point that concerns decoupling of doublers. On
this extreme strong coupling ($g_2\gg 1$) symmetric segment, the $\psi^i_L$ and
$\psi_R$ in (\ref{action}) are bound up to form the three-fermion
states\cite{ep,xue96}: 
\begin{equation}
\Psi_R^i={(\bar\psi_R\cdot\psi^i_L)\psi_R\over
2a};\hskip0.1cm\Psi^n_L={(\bar\psi_L^i
\cdot\psi_R)\psi_L^i\over 2a},
\label{bound}
\end{equation}
which carry the appropriate quantum
numbers of the chiral group that accommodates $\psi^i_L$ and
$\psi_R$. These three-fermion states are Weyl
fermions and respectively pair up with the $\bar\psi_R$ and $\bar\psi_L^i$
to be massive, neutral $\Psi_n$ and charged $\Psi_c^i$
Dirac fermions, 
\begin{equation}
\Psi^i_c=(\psi_L^i, \Psi^i_R),\hskip0.3cm\Psi_n=(\Psi_L^n, \psi_R).
\label{di}
\end{equation}
To show this phenomenon, we compute the following two-point functions,
\begin{eqnarray}
S^{ij}_{LL}(x)&=&
\langle\psi^i_L(0)\bar\psi^j_L(x)\rangle,\nonumber\\
S^{ij}_{ML}(x)&=&\langle\psi^i_L(0)\bar\Psi^j_R(x)\rangle,\nonumber\\
S^{ij}_{MM}(x)&=&\langle\Psi^i_R(0)\bar\Psi^j_R(x)\rangle.
\label{twopoint}
\end{eqnarray}
In the lowest non-trivial order $O({1\over g_2})$, we obtain the
following recursion relations 
\begin{eqnarray} 
S^{ij}_{LL}(x)\!&\!=\!&\!{\sum^\dagger_\mu S^{ij}_{ML}(x\!+\!\mu)\gamma_\mu
\over 8a^3g_2\Delta^2(x)},\label{re11}\\
S^{ij}_{MM}(x)\!&\!=\!&\!{\sum^\dagger_\mu 
\gamma_\mu\gamma_\circ S^{ij\dagger}_{ML}(x\!+\!\mu)\gamma_\circ
\over 8a^3g_2\Delta^2(x)},
\label{re31}\\
S^{ij}_{ML}(x)\!&\!=\!&\!{\delta(x)\delta_{ij}\over 4ag_2\Delta^2(x)}\nonumber\\
&\!+\!&{\sum^\dagger_\mu S^{ij}_{LL}(x\!+\!\mu)\gamma_\mu\over 8a^2g_2\Delta^2(x)}
.\label{re21}
\end{eqnarray}
Eq.(\ref{re21}) suggestes that the states coupling to operators
$\psi^i_L(x)$ and $\Psi^i_R(x)$ are mixed, producing a massive 
four-component Dirac 
fermion. To find the masses, we make the Fourier transform of these recursion
equations for $p\not=0$ and $\Delta^2(p)=4w^2(p)\not=0$ and obtain
the inverse propagators of the charged Dirac fermions, 
\begin{equation}
S_c^{-1}(p)={i\over a}\sum_\mu\gamma_\mu \sin p^\mu 
+8ag_2w^2(p).
\label{sc1}
\end{equation}
This shows that all doublers $p=\tilde p+\pi_A$
are decoupled as massive Dirac fermions consistently with the
$SU_L(2)\otimes U_R(1)$ chiral symmetries. 

We are left with the last point that undoubled chiral fermions $(p=\tilde
p\sim 0)$ of $\psi^i_L$ exist in this
segment. We repeat the same calculations and 
discussions in 1+1 dimensions, where the time direction is continuous and 
space is discrete. We obtain the dispersion relation corresponding to the 
Dirac fermion (\ref{sc1}) for $P\not=0$,
\begin{equation}
E(P)=\pm\sqrt{\sin^2P+(8a^2gw^2(P))^2},
\label{des}
\end{equation}
which is analytically continuous to low momentum states $P\rightarrow 0$,
unless it hits the threshold. For a given total momentum $P$, we consider the
system that contains three free chiral fermions, i.e.~right-movers $\bar\psi_R,
\psi_R$ with momenta $p_1$ and $p_2$ and a left-mover $\psi_L^i$ with momentum
$p_3$ ($|p_i|\ll {\pi\over2}, i=1,2,3$), 
\begin{equation}
P=p_1+p_2+p_3\ll {\pi\over2}.
\label{totalm}
\end{equation}
Since NJL spontaneous symmetry breaking does not occur when 
$|p_i|\rightarrow 0$, the total energy of such a system is given by
\begin{eqnarray}
E&=&E_1(p_1)+E_2(p_2)+E_3(p_3)\nonumber\\
E_i&=&\pm p_i, 
\label{totale}
\end{eqnarray}
where all negative-energy states have been filled. 
The lowest energy (the threshold) of such a system and corresponding 
configuration are 
\begin{eqnarray}
E_t(P)&=&3\sin P,\label{thre}\\
p_3&=&-P, p_1=p_2=P,\nonumber 
\end{eqnarray}
The three-fermion state is stable if there is a gap $\delta$ between
the threshold (\ref{thre}) and the energy (\ref{des}) of the three-fermion 
state, i.e.
\begin{equation}
\delta=E_t(P)-E(P)>0.
\label{sta}
\end{equation}
The three-fermion state disappears, when the gap $\delta$ goes to zero,
\begin{equation}
\delta=E_t(P)-E(P)=0.
\label{con}
\end{equation}
Substituting eqs.(\ref{des}) and (\ref{thre}) into eq.(\ref{con}), we obtain
in the continuum limit $P\rightarrow 0$, the gap
\begin{equation}
\delta(P)\rightarrow 0,
\label{fin}
\end{equation}
where the three-fermion-state spectrum
dissolves into free chiral fermion spectra. As a result, an undoubled
chiral fermion exists in the continuum limit.

I thanks Profs.~G.~Preparata, H.~B.~Nielsen and E.~Eichten for discussions
on the last point.

\end{document}